\documentclass[twocolumn]{aastex631}

\usepackage{newtxtext,newtxmath}
\usepackage{physics} 
\usepackage{amssymb} 
\newcommand\mbf[1]{\mathbf{#1}}
\usepackage{mathtools}

\begin{document}

\title{Ion-electron instabilities in the precursor of weakly magnetized, transrelativistic shocks}

\author[0000-0002-0790-5481]{Théo Abounnasr}
\affiliation{Instituto de Plasmas e Fusão Nuclear, Instituto Superior Técnico, Universidade de Lisboa, 1049-001 Lisboa, Portugal}
\email{tabounnasrmartins@ipfn.ist.utl.pt}




\begin{abstract}
Collisionless shocks are known to induce turbulence via upstream ion reflection within the precursor region. This study elucidates the properties of electrostatic and electromagnetic instabilities, exploring their role over the transrelativistic range. Notably, the growth of oblique Buneman waves and previously overlooked, secondary small-scale ion-electron instabilities is observed to be particularly promoted in the relativistic regime. Furthermore, the growth of large-scale electromagnetic modes encounters severe limitations in any cold, baryon-loaded precursor. It is argued that electron pre-heating in electrostatic modes may alleviate these constraints, facilitating the subsequent growth of large-scale filamentation modes in the precursor of transrelativistic shocks.
\end{abstract}



\section{Introduction} \label{sec:intro}

As a promising site of acceleration for non-thermal populations of particles, shock waves in plasmas have focused considerable theoretical efforts. Shock acceleration models have found some success in explaining the low-energy cosmic-ray spectrum with non-relativistic shocks e.g interplanetary shocks \citep{Zank2000} and Supernovae Remnants (SNR) \citep{Bell2013}, although not without tensions \citep{malkov2024flat,Hillas2005}. 
Turning to extra-galactic candidates for the acceleration of ultra-high-energy cosmic-rays (UHECR), gamma-ray burst (GRB) events are a case of prime interest as they typically involve a variety of transrelativistic shocks. The afterglow emission is commonly associated with unmagnetized ultra-relativistic blast waves with Lorentz factors $\Gamma_{sh} \gtrsim \mathcal{O}(100)$ propagating in the interstellar medium (ISM) \citep{Piran2005}. The detection of GRB170817 following the gravitational-wave signal of a binary merger \citep{Abbott_2017} has prompted further consideration into mildly relativistic shocks associated with a kilonova emission \citep{Smartt2017} or arising from a cocoon-like structure  $\Gamma_{sh} = \mathcal{O}(1\sim10)$ \citep{Kasliwal2017,Gottlieb2018}.
\\
The shock formation is naturally associated with plasma instabilities \citep{Bret2013}, which are responsible for the development of turbulence in the \textit{precursor}. This turbulent environment is a necessary condition to allow the successive scattering of particles from either side of the shock in the Diffusive Shock Acceleration (DSA) model. Within this picture, instabilities mediate the shock's structure and dissipation of energy into accelerated particles.\\ Extrapolating from in-situ observations of Earth’s bow shock \citep{Paschmann1980}, it is expected that instabilities arise in the plasma from the interaction with a returning beam of ions reflected at the shock front \citep{Leroy1981}. It is in this context that \cite{Shimada2000} confirmed the operation of a mechanism first proposed by  \cite{Papadopoulos1971}, whereby electrons are heated in electrostatic waves excited by the ``Buneman'' instability \citep{Buneman1958} (hereafter BI). Recent simulations \citep{Vanthieghem2022} confirmed similar mechanisms at relativistic shock waves, with electron heating however taking place in a precursor dominated by electromagnetic turbulence. This development can be attributed to the ``Filamentation Instability'' (FI) \citep{Fried1959,Weibel1959}\footnote{This terminology is preferred to ``Weibel Instability'', following a remark of \cite{Bret2004} on the potentially confusing reference to another class of instability.}, also associated with the growth of transverse magnetostatic fluctuations, that has long been speculated to mediate the relativistic shocks of GRBs \citep{Medvedev1999}. 
Whether it is the case, or not, is an especially pressing question since magnetic turbulence has been shown to play a major role in operating DSA at relativistic shocks. It has the potential of reviving the process at superluminal configurations where it could not otherwise operate \citep{Begelman1990,Bresci2023} while also significantly affecting the spectrum of accelerated particles \citep{Niemiec2004}. Understanding the conditions under which shocks are unstable to the FI could thus uncover GRBs as credible counterparts for the acceleration UHECR.
\\~\\
It was discussed by \cite{Lyubarsky2006} (hereafter LEO6) that angular dispersion of the beam would quench the instability irremediably. \cite{Lemoine2011} (LP11) however demonstrated that it could eventually be recovered by aberration effects in the ultra-relativistic limit $\Gamma_{sh} \gtrsim \mathcal{O}(100)$. In this regime, additional limitation due to shrinking precursor size must be considered for a self-consistent treatment, see also \cite{Rabinak2011}. In LP11 and \cite{Plotnikov2013} was also noted the potential role of oblique electrostatic waves in pre-heating the precursor to temperatures favourable for the growth of large-scale electromagnetic modes.  While the existence of these oblique modes has been merely evoked to sustain isotropic heating observed in the linear stage of non-relativistic simulations \citep{Amano2009}, further attention is due as their growth is expected to be promoted in the relativistic regime (\cite{Bret2004}, B04).
\\
In this context, it appears fundamental to identify the transition thoroughly examining the turbulent content generated as a function of $\Gamma_{sh}$ with the aim of eventually identify a transition from electrostatic to electromagnetic instabilities. To assess the stability of beam-plasma systems over the complete transrelativistic range, linear theory appears to be a particularly suitable approach. This work thus aims to provide a unified description of electromagnetic instabilities in the precursor of transrelativistic shocks. The general linear theory of electrostatic instabilities is presented in \autoref{sec:ES}, with electron kinetic effects discussed in Appendix \ref{sec:A_TS}. The discussion extends to perpendicular electromagnetic modes in \autoref{sec_EM}, addressing inconsistencies in the literature and unveiling a secondary FI. Results are compiled in \autoref{sec:transre}, showing a double limitation on the growth of large-scale filamentation modes as a function of shock Lorentz factor. Finally, a summary and a discussion of the results are presented in \autoref{sec:summary}, before concluding in \autoref{sec:conclusion}.
\section{Electrostatic instabilities}
\label{sec:ES}
The dispersion of electrostatic perturbations $\mbf{k}\cross \mbf{E}=\mbf{0}$ is obtained from the roots of a single dielectric scalar  $\varepsilon_{L} \equiv 1+\sum_s \chi_s$ \citep{Davidson_1983}, with
\begin{equation}
    \chi_s=-\omega_{ps}^2\int \dd^3 \mbf{u}~\frac{f_s(\mbf{u})}{\qty(\omega - \mbf{k}\cdot\mbf{u}/\sqrt{1+u^2})^2}\label{Cold_dielec_es}
\end{equation}
the susceptibility of a given species `s' in the plasma. For a cold relativistic population represented by a dirac-delta distribution of four-velocities $f_s(\mbf{u})=n_s\delta(u-u_s)\delta^{(2)}(\mbf{u_\perp})$, it is written \begin{equation}
    \chi_s=-\frac{\omega_{ps}^2}{\gamma_s k^2}\frac{k_x^2+k_z^2/\gamma_{s}^2}{\qty(\omega - k_zV_s)^2}. \label{eq:chi_s}
\end{equation}
The Lorentz boost along the beam is reflected in the different scalings for the parallel and perpendicular projection of the wavenumber, $k_z$ and $k_x$ respectively. The dielectric response of each species is further characterized by the plasma frequency $\omega_{ps}$, Lorentz factor $\gamma_s=\sqrt{1+u_s^2}$ and corresponding drift 3-velocity $V_s=u_s/\gamma_s$. The non-relativistic limit $\gamma_s\simeq1$ recovers a dielectric scalar independent on the perpendicular wave-number $k_x$, as obtained in standard frameworks considering non-relativistic plasmas (e.g \citet{Galeev1983}). 
\\~\\
The interactions within a three-population plasma composed of an ion beam `b' and a background of electrons `e' and ions `i' allows for the propagation of electrostatic waves verifying the dispersion relation 
 \begin{equation}
    \omega_{pe}^{-2} - \frac{1}{\gamma_e k^2}\frac{k_x^2+k_z^2/\gamma_{e}^2}{\qty(\omega - k_zV_e)^2}  - \frac{\alpha_b/R_b}{\gamma_b k^2}\frac{k_x^2+k_z^2/\gamma_{b}^2}{\qty(\omega - k_zV_b)^2}-\frac{\alpha_i/R_i}{\omega^2}=0. \label{eq:bunemanfull_2D}
\end{equation}
Here the parameters for ion populations $s=\{b,i\}$, $\alpha_s \equiv (\gamma_b/\gamma_e)(n_{s0}/n_{e0})$ and $R_s=(\gamma_s/\gamma_e)(m_{s0}/m_{e0})$, represent the upstream specie 's' to electron 'e' density and mass ratios, respectively. The ratio of plasma frequencies $(\omega_{ps}/\omega_{pe})^2=\alpha_s/ R_s$ does yield a Lorentz-invariant quantity nonetheless. \\
In \eqref{eq:bunemanfull_2D}, the susceptibility of all species are written in the cold limit with \eqref{eq:chi_s}, while kinetic effects will be considered later with \eqref{eq:bunneman_hot1D}. In any case, a finite drift is considered for electrons $V_e=\alpha_b V_b$ to neutralize the current of the beam \citep{Bret2004}. The other contributions will remain unchanged throughout this section. The second term is thus specified representing a cold, relativistic beam streaming with velocity $V_b$ and density $n_b=\alpha_b n_e$. The third term accounts for the presence of static ions in their inertial frame, $V_i=0$. 

\begin{figure}
    \centering
    \includegraphics[width=\linewidth]{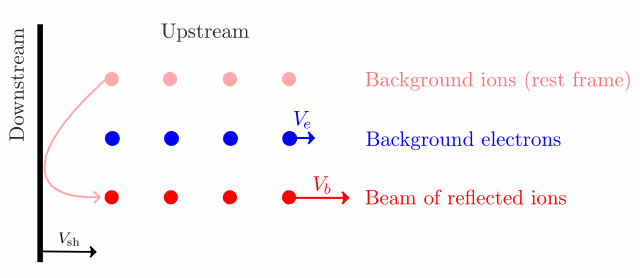}
    \caption{Sketch of the three-population setup in the rest frame of background ions, as considered for equation \eqref{eq:bunemanfull_2D}. In this frame, the shock is moving with a velocity $V_{sh}$ towards the unshocked medium (upstream). The thick black line represents the shock discontinuity in which the background ions assumed to be reflected. Electrons are represented with finite drift velocity so as to neutralize the current of the ion beam streaming upstream.}
    \label{fig:sketch_beam}
\end{figure}

\subsection{Principal Buneman instability in cold plasma}
In the cold limit, the susceptibility of electrons in written with \eqref{eq:chi_s}. Considering waves in the vicinity of $\omega \simeq k_z V_b$, a hierarchy $\chi_b > \chi_e \gg \chi_i$ can be established in virtue of $\chi_i/\chi_e \propto \qty(\omega_{pi}/\omega_{pe})^2=1/R_i \ll 1$. The interaction of massive background ions being thus neglected compared to the fast response of electrons, the solutions to \eqref{eq:bunemanfull_2D} correspond to a two-stream-like interaction between beam ions and background electrons. It follows that Langmuir waves propagating in \textit{Cerenkov} resonance with the beam, $k_z V_b=\omega_{pe}$, are susceptible to the Buneman instability with the dominant $\mathrm{B_1}$-mode growing at a rate (see also \citet{Nakar_2011}):\begin{equation}
    \Psi_{\mathrm{B_1}}^{\textrm{cold}}=\frac{\sqrt{3}}{\gamma_b}\qty(\dfrac{\alpha_b}{16R_b}\left[\cos^2\theta + \gamma_b^2\sin^2\theta\right])^{1/3}.\label{eq:BI_growth}
\end{equation}
The Lorentz factor dependence is a function of the angle $\theta\equiv \arctan k_x/k_z$ at which the  perturbation propagates with respect to the beam-axis $\hat{z}$. The electrostatic model presented here allows to generalize derivations presented in \cite{Bludman1960} for the two-stream instability ($\theta=0$) to arbitrary orientations of the wave-vector, by avoiding the otherwise electromagnetic features appearing at finite $k_x$. The formula \eqref{eq:BI_growth} thus provides an analytical basis for the dominance of oblique modes in the relativistic regime as pointed out in the electromagnetic model of \cite{Bret2004}, noting that $\Psi_{\mathrm{B_1},90^{\circ}}/\Psi_{\mathrm{B_1},0^{\circ}}=\gamma_b^{-2/3}$.\\~\\  
This effect, already manifest in the susceptibility \eqref{eq:chi_s}, can be attributed to the boost of an electric field perturbation perpendicular to the parallel component of the Buneman wave co-moving with the beam. It means that electrons in the upstream frame would experience an effectively reduced inertia when moving along the beam-axis, resulting in the modulation of the growth rate by a factor $\gamma_b^{-2/3}$ for $k_z \gg k_x$ waves. 

\subsection{Secondary Buneman instability in a cold plasma}

In fact, the finite drift velocity of electrons required to neutralize the current of the beam can trigger another Buneman instability. This implies looking for a solution in the vicinity of the electron drift resonance $\omega \simeq k_z V_e$. In this limit, the beam resonance term can be overlooked in virtue of the hierarchy $\chi_e \gg \chi_i > \chi_b $, which holds when the $\mathrm{B_2}$ resonance is sufficiently separated from the $\mathrm{B_1}$ resonance i.e as long as $\qty(\omega_{pb}/\omega_{pi})^2= n_b/n_i < 1$ is true. Buneman unstable "$\mathrm{B_2}$-modes", are thus understood to arise from the interaction of drifting electrons with background ions triggering unstable Langmuir waves at the electron drift $\mathrm{B_2}$ resonance $\Re{\omega} \simeq k_z V_e=\alpha_b k_z V_b$, with maximum growth rate \begin{equation}
    \Psi_{\mathrm{B_2}}^{\textrm{cold}}=\frac{\sqrt{3}}{(16R)^{1/3}} \simeq \qty(\frac{m_e}{m_i})^{1/3}. \label{eq:growth_BII}
\end{equation}
This formula holds as long as $\alpha_b \ll 1$. As the electron response becomes relativistic relativistic, the expression \eqref{eq:BI_growth} must be used substituting $\gamma_b \leftarrow \gamma_e$, $\alpha_b \leftarrow \alpha_i \equiv n_i/n_e$ and $R_b \leftarrow R_i \equiv m_i/m_e$. The potential role of these modes, widely overlooked in the literature --- only a mention found in \cite{Ohira2008} --- is discussed in \autoref{sec:ES_heating}. 
\subsection{Buneman instability in a hot electrons background}
\begin{figure}
	\includegraphics[width=\columnwidth]{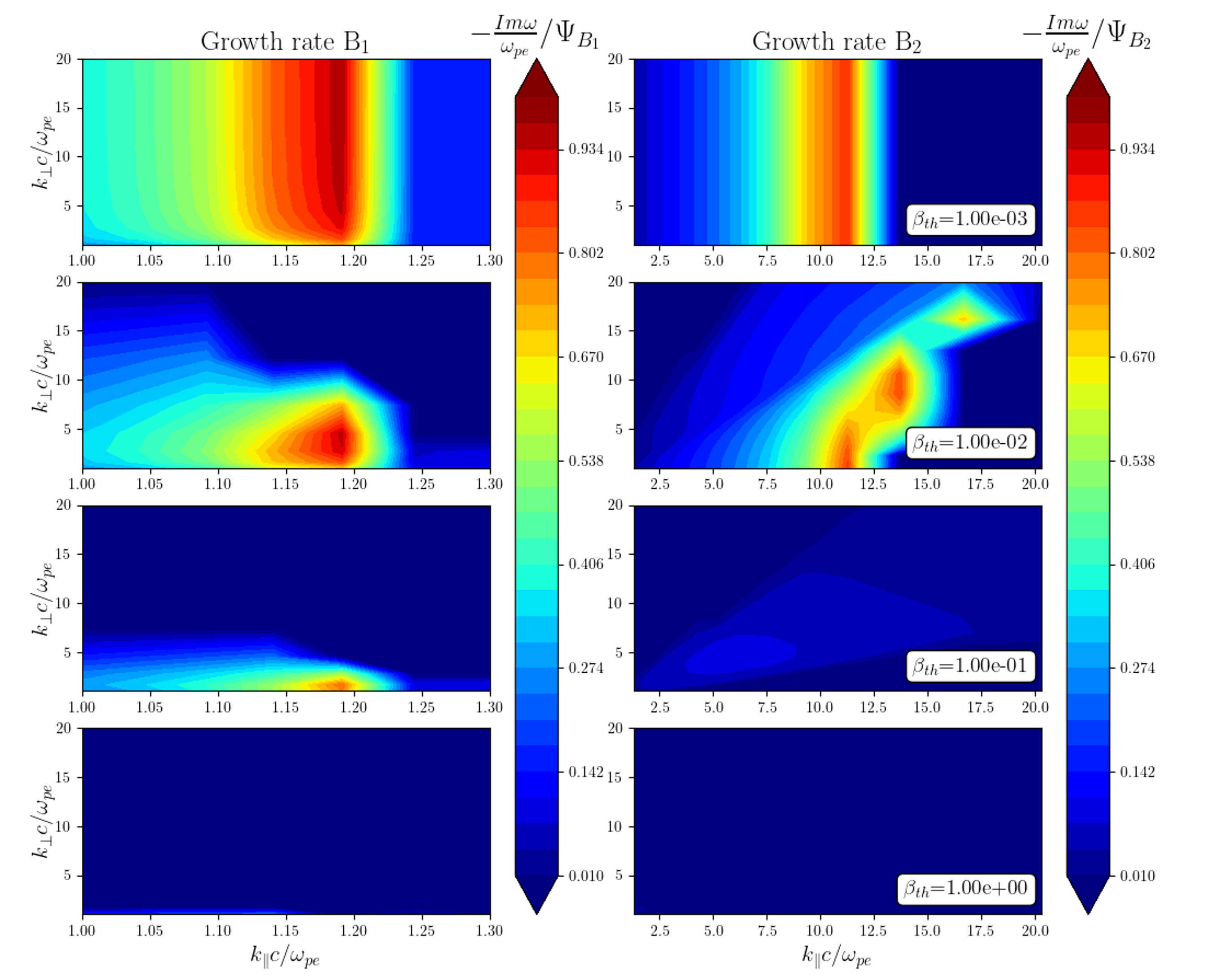}
    \caption{Two-dimensional spectrum for the growth rate of Buneman-type instabilities at $\alpha_b=0.1$, $\gamma_b=3$, $R_b=1836$. Plots for $\mathrm{B_1}$ and $\mathrm{B_2}$ modes are separated for visualization purpose but belong to the same wavenumber space. For the chosen set of parameters, left resp. right colors are normalized by the maximum value $\Psi_{\mathrm{B_1}}=0.018\omega_{pe}$ resp. $\Psi_{\mathrm{B_2}}=0.056\omega_{pe}$.  The $\mathrm{B_2}$ spectrum is also plotted over a much wider range of $k_\parallel$ values, thus exhibiting a much broader resonance than the $\mathrm{B_1}$-resonance as well as a faster growth rate. The obliquity of the $\mathrm{B_2}$-resonance at finite temperature is a relativistic feature analogous to what is observed with a Maxwellian distribution in \cite{Nakar_2011}. A notable difference being the use of the waterbag approximation in this study, thus capturing accurately only the first harmonic of a series of otherwise infinite oblique modes.}
    \label{fig:BIandBII_3}
\end{figure}

To evaluate electron temperature effects in the plasma analytically, a standard approach in the literature is to approximate the thermal spread of a Maxwellian distribution with a ``Waterbag'' function (e.g \cite{Yoon_87,Gedalin_99,Bret2004}),
\begin{equation}
    W_{th}(u)=\frac{1}{2u_{th}}\qty(\Theta(u-{u_{th}})-\Theta(u+{u_{th}})). \label{eq:waterbag_3D}
\end{equation}
The simplest distribution accounting for electron thermal dispersion $u_{th}$ and drift $u_e$ is $f_e(\mathbf{u})=n_eW_{th}(u_\parallel-u_e)W^{(2)}_{th}(\mbf{u_\perp})$. The dispersion relation over the whole spectrum of $(k_z,k_x)$ is solved at different temperatures to obtain the growth rates shown is \autoref{fig:BIandBII_3}. It can be observed that electron temperature progressively quench the growth of finite $k_x$ modes, until all modes are stabilized all at once as electrons reach a temperature $u_{th} \sim u_b$. The parallel modes $k_x=0$ being the last to be unstable, the effects of temperature will be thereafter analyzed in this limit only. The kinetic effects in the one-dimensional limit modifies the second term in the dispersion relation \eqref{eq:bunemanfull_2D} to a tractable analytical expression, \begin{equation}
    \chi_{e}=-\frac{\omega_{pe}^2}{2k_z c u_{th}}\qty(\frac{1}{\omega-k_z c\beta+}-\frac{1}{\omega-k_z c\beta_-}). \label{eq:bunneman_hot1D}
\end{equation}   
The expression is kept covariant by summing relativistic velocities according to the transformation \begin{equation}
    \beta_{\pm} \equiv  \frac{u_e \pm u_{th}}{(1+(u_e\pm u_{th})^2)^{1/2} } \underset{\ll 1}{\simeq} \beta_e \pm \beta_{th}. \label{eq:bpm}
\end{equation}

\subsubsection{$\mathrm{B_1}$ in a hot plasma}
Neglecting once again the background ion term near the cerenkov resonance $\omega \simeq k_z V_b$ mode, the electron kinetic effects are investigated. The derivation procedure to solve the dispersion relation including \eqref{eq:bunneman_hot1D} as the electron term is presented in Appendix \ref{sec:A_TS}. The maximum growth rate is obtained as \begin{equation}
    \Psi_{\mathrm{B_1}}(u_{th})=\Psi_{\mathrm{B_1}}^{\textrm{cold}} \qty(\frac{\qty(\beta_b-\beta_{-})^{-2}-(\beta_b-\beta_{+})^{-2}}{4u_{th}})^{1/3},~\beta_b \neq \beta_\pm \label{eq:growth_BI_hot}
\end{equation} 
The kinetic effects are decoupled from the magnitude of the cold growth rate. It can be indeed verified that the factor in parenthesis converges to unity in the cold limit $u_{th} \rightarrow 0$.\\~\\
Although a stabilization criterion does not appear obviously from \eqref{eq:growth_BI_hot}, it is possible to use \eqref{eq:A_thresh} instead and derive a threshold wave-number \begin{equation}
    k_c=\frac{\omega_{pe}}{c}\qty(2u_{th}\frac{\qty(\beta_b-\beta_{+})\qty(\beta_b-\beta_{-})}{\beta_+-\beta_-})^{-1/2}.
\end{equation}
This expression converges to $k_c=\frac{\omega_{pe}}{c}(1+\frac{3}{2}\frac{(\alpha_b/R_b)^{1/3}}{\gamma_b})$ in the cold and $\alpha_b \ll 1$ limit, consistently with results obtained in B04. The shift toward higher wave-numbers numbers with increasing temperature as seen on \autoref{fig:BIandBII_3} is also observed. This threshold wave-number is zero in the limit $\beta_+ \rightarrow \beta_b$, which means that no imaginary solutions are allowed beyond this limit and the stability criterion for the Buneman instability in a hot plasma can be written \begin{equation}
    u_{th} \geq  u_b-u_e, \label{eq:Penrose_BI}
\end{equation}
which is a natural extension of the so called "Penrose criterion" $v_{th} \geq v_b$ in the non-relativistic limit.
\subsubsection{$\mathrm{B_2}$ in a hot plasma}
Now neglecting the beam term near the $\mathrm{B_2}$-resonance $\omega_{pe}\simeq k_z V_e$, the derivation for the growth rate is easily reproduced for the secondary instability, to find 
\begin{equation}
    \Psi_{\mathrm{B_2}}(u_{th})=\Psi_{\mathrm{B_2}}^{\textrm{cold}} \qty(\frac{\beta_{-}^{-2}-\beta_{+}^{-2}}{4u_{th}})^{1/3},~\beta_e \neq \beta_\pm,\label{eq:growth_BII_hot}
\end{equation} 
and the threshold wave-number is 
    \begin{equation}
    k_c=\frac{\omega_{pe}}{c}\qty(2u_{th}\frac{\beta_{+}\beta_{-}}{\beta_+-\beta_-})^{-1/2}. \label{eq:thres_BII}
\end{equation}
Increasing the temperature, $k_c$ reaches zero when $\beta_-=0$, so that the stabilization criterion for $\mathrm{B_2}$ modes can be written
\begin{equation}
    u_{th} \geq u_e, \label{eq:criterion_BII}
\end{equation}
 
\subsection{Implications for electron heating}
\label{sec:ES_heating}
\subsubsection{Heating in $\mathrm{B_1}$ modes}
The interactions of beam ions with background electrons trigger $\mathrm{B_1}$ waves, the most unstable of which travel along the beam-axis with a phase velocity \begin{equation}
    v_\phi \sim V_b.
\end{equation}
Considering that an ion beam is continuously reflected away from the shock, the extent of which defines the length of the precursor, the growth of $\mathrm{B_1}$-unstable waves can be expected to take part in heating within that region. As electrons are progressively heated up to temperatures prescribed by \eqref{eq:Penrose_BI}, the plasma near the shock-transition becomes Buneman-stable and $\mathrm{B_1}$ modes can only survive at the tip of the precursor. Although a robust understanding of the physical mechanisms involved in this scenario is still out of reach of the current state of the art, insights from numerical studies can shed a light on the dynamics beyond linear theory. While the linear stage would display a constant growth rate \eqref{eq:BI_growth}, the dynamics of the instability becomes non-linear as mode-coupling interactions eventually quench the growth rate. For non-relativistic flows, early numerical investigations \citep{Lampe1974, Ishihara1981} of 1D Buneman instability dynamics prescribe a saturated field energy of the order of initial drift kinetic energy. As the not-yet-undertaken development of a relativistic framework is far beyond the scope of this paper, we simply suggest that once the field energy is of the order of relativistic kinetic energy \begin{equation}
    \frac{E^2}{8\pi}\simeq (\gamma_b-1)n_b m_e c^2, \label{eq:rel_crit}
\end{equation}
$\mathrm{B_1}$ waves become capable of trapping electrons, leading to the saturation of the instability. Besides, it is known from the same simulations studies that the turbulent field energy constitutes the reservoir for electron thermal energy $(\gamma_{th}-1) m_e c^2$. It would follow that electron are heated to relativistic temperatures $\gamma_{th}\simeq \gamma_b$ at saturation. This energetic argument matches the stabilization condition \eqref{eq:Penrose_BI}, derived from the  kinetic damping of the linear growth rate. 
\\~\\
A significantly lower saturation level has been observed in non-relativistic 2D Buneman instability PIC simulations \citep{Amano2009}, attributed to the early saturation of modes propagating perpendicular to the beam. The promotion of oblique modes in the relativistic regime is then understood to modify the usual picture in two regards.
(i) Changes in the non-linear dynamics may be reflected into significant changes of the turbulent field saturation level \eqref{eq:rel_crit}; (ii) The linear stage of electron heating in electrostatic modes is now expected to be anisotropic. This will in turn have some consequences for the further development of electromagnetic instabilities, discussed in \autoref{sec_EM}.

\section{Electromagnetic instabilities}
\label{sec_EM}
When perturbations oblique to the electric field are allowed, the dielectric response is represented by a tensor \cite{ichimaru1973basic}
\begin{equation}
\begin{split}
    \varepsilon_{ij} = \delta_{ij} &+ \sum_s \frac{\omega_{ps}^2}{n_s \omega^2}\int \dd^3\mbf{u}~ \frac{u_i}{\gamma(\mbf{u})^2}\pdv{f_0}{u_j} \\&+  \sum_s \frac{\omega_{ps}^2}{n_s \omega^2}\int \dd^3 \mbf{u}~\frac{u_i u_j}{\gamma(\mbf{u})^2} \frac{\mbf{k} \cdot \pdv{f_0}{\mbf{u}}}{(\omega+i/\tau) - \mbf{k}\cdot\mbf{v}(\mbf{u})}.
    \label{eq:dielectric_tensor}
\end{split}   
\end{equation}~\\
It follows that electromagnetic waves $\mbf{k}\cross\mbf{E}\neq 0$ with frequencies $\omega<\omega_{pe}$ can penetrate in the plasma up to a skin depth $l_e=(\omega_{pe}/c)^{-1}$. Charge separation is most efficient in directions transverse to the beam, and can lead to low-frequency, large-scale electromagnetic modes susceptible to the Filamentation Instability, coined `FI' thereafter. These modes couple to local density perturbations, inducing induce both a Direct Current ($\Re{\omega} = 0$)  potential along the beam axis and secular magnetic field turbulence transverse to the beam-perturbation plane on $k^{-1}_x$ scale. The persistent filamentary structure surrounding localized current channels is well documented by PIC simulations as well as laboratory experiments (e.g \cite{Silva_2003,Hededal_2004,Allen_2012}).
\subsection{Cold limit}
The calculations of the susceptibility tensor's components have been the object of previous studies \citep{Bret2004,Lemoine2011} and are here reported in Appendix \ref{sec:A_FI}. The most unstable modes being found in $k_z=0$, the dispersion relation will be calculated in this limit. In the cold case, this implies a single resonance in $\Re{\omega}=0$ independently on the number of species considered in the plasma. The ratio of plasma frequency will thus prescribe a hierarchy $\hat{\chi}_e > \hat{\chi}_b > \hat{\chi}_i$, from which it follows that the response of background ions can be neglected in the cold approximation. The dispersion relation of electromagnetic waves propagating in a cold two-population plasma is given by the zeros of the function

\begin{widetext}
\centering
\begin{equation}
 F_{k_x}(\omega)=\qty(\frac{\omega^2}{\omega_{pb}^2} -\frac{1}{\gamma_b} - \frac{R_b}{\alpha_b \gamma_e})\qty(\frac{\omega^2}{\omega_{pb}^2} -\frac{1+k_x^2u_b^2/\omega^2}{\gamma_b^3}-\frac{R_b(1+k_x^2u_e^2/\omega^2)  }{\alpha \gamma_e^3} -  \frac{k_x^2 c^2}{\omega_{pb}^2}) - \frac{k_x^2}{\omega^2}\qty(\frac{R_b}{\alpha}\frac{u_e}{\gamma_e^{3/2}}+\frac{u_b}{\gamma_b^{3/2}})^2.
\end{equation}
\end{widetext}
Here, the response of electrons is kept relativistic for the purpose of qualitative discussion further ahead. Setting $\Re{\omega}=0$, the dispersion relation can be solved exactly to find a growth rate 
\begin{align}
    &\Im{\omega}(k_x)=\Psi_{\mathrm{F_1}}\omega_{pb}\frac{\abs{k_x}}{\sqrt{k_x^2+(\omega_{pe}/c)^2}},\\    
    &\Psi_{\mathrm{F_1}}=\frac{\beta_b}{\sqrt{\gamma_b}}\gamma_e\qty(1-\alpha_b)\qty(1+\frac{\alpha_b\gamma_e^3}{R_b\gamma_b^3})^{-1/2}. \label{eq:growth_FI_cold}
\end{align}
The maximum growth rate, found in the limit $k_x\rightarrow\infty$, is given by $\Psi_{\mathrm{F_1}}$ \eqref{eq:growth_FI_cold}. Aforementioned references have reported this result, without the corrections in $\alpha_b$, $\gamma_e$. The factor $1-\alpha_b$ here is of great importance as it recovers a stable plasma in the limit $\alpha_b \rightarrow 1$ where the beam and electrons move together as a single neutral flow. This highlights the limitations of a two-population model as the background ion population can not be neglected.\\~\\
Considering the interactions within the background plasma arising from the finite drift velocity of electrons in the precursor frame, a secondary FI, coined `$\mathrm{F_2}$', is identified with growth rate \begin{equation}
    \Psi_{\mathrm{F_2}}\omega_{pe} \simeq \frac{\beta_e}{\sqrt{\gamma_e}} \omega_{pi}
\end{equation}
Approximateively, it can be shown that the beam-driven instability $\mathrm{F_1}$ is dominant as long as $\alpha_b \gamma_b \ll 1$. As the beam gets \textit{stronger}, the electrons are dragged to neutralize the currents more efficiently and the relative velocity between electrons and background ions become smaller than with the beam. In the electron frame, it would then appear that background ions now act as the main beam, and therefore drive the instability. This transition between the two ion populations as driver will prove to be of key importance when considering kinetic effects.
\subsection{Kinetic effects}
\label{sec:FI_kin}
The analysis of kinetic effects attributed to both electron temperature and beam velocity spread on the FI are particularly relevant when considered together with beam velocity spread\footnote{Here the distinction is essentially conceptual. The electron \textit{thermal spread} is considered as a dynamical parameter associated with heating in turbulence, typically the second moment of a Maxwellian distribution, whereas the beam \textit{velocity spread} is considered as an initial parameter accounting for the dispersion in the beam of reflected ions. For all practical matters, the definitions are equivalent in this framework and will be used interchangeably.}. 
The waterbag distibution \eqref{eq:waterbag_3D} employed as an approximation for electron temperature also happens to accurately represent the truncated, uniform dispersion of velocities in the beam obtained from the reflection model detailed in Appendix \ref{sec:A_K}. On the one hand, the model predicts a beam temperature mostly anisotropic over the transrelativistic range, with the ratio $\rho_{b\parallel} \equiv \Delta u_\parallel/u_b$ monotonically decreasing from unity to $\sim 1/3$ in the ultra-relativistic limit \eqref{eq:rhob}. On the other hand, terms involving $\rho_{b\parallel}$ are negligible in the dispersion relation as long as $\rho_{b\parallel} \ll 1$. Therefore parallel temperature is neglected altogether and \[f_b=n_b\delta(u-u_b)W^{(2)}[{\Delta u_\perp}](\mbf{u_\perp}) \] is a sufficient kinetic model for the distribution of the beam. Introducing a single parameter $\rho_b= \rho_{b\perp}\equiv \Delta u_\perp/u_b$  allows to simplify the analysis considerably. While the beam dispersion is naturally motivated and constrained by the reflection model ($\rho_b  \propto 1/\Gamma_{sh}$), it is kept a free parameter in this section to highlight its role in the properties of the FI.

\subsubsection{Problems of two-population models}
 Previous literature has exclusively focused on two-population models, systematically ignoring the presence of background ions on the spectrum of unstable modes. The updated dispersion relation is reported in Appendix \ref{sec:A_FI}. On \autoref{fig:Spectrum_FI}, dashed lines represent the growth rate of numerical solutions to the two-population dispersion relation. For these, it can be seen that beam velocity spread brings back the spectrum of $\mathrm{F_1}$ modes to larger scales according to the ratio of four-velocity spread to average, $\rho_b \equiv \Delta_\perp u_b /u_b$. Disregarding background ions, the last unstable wavenumber $k_c$ is solution to $F_{{k_c}}(\Re{\omega}=0)=0$, that is \begin{equation}
    \frac{k_c c}{\omega_{pb}}= \frac{\gamma_e(1-\alpha_b)}{\rho_{b \perp}\sqrt{\gamma_b}} \qty[1-\gamma_b\rho_{b\perp}^2\frac{R_b}{\alpha_b\gamma_e^3}\qty(1+\qty(\frac{\gamma_e}{\gamma_b})^3 \frac{\alpha_b}{R_b})]. \label{eq:F_treshc}
\end{equation}
The stabilization condition in a cold plasma can then be written as uncovered by LE06, neglecting higher-than-first order $\alpha_b$ terms: \begin{equation}
    \rho_b \gtrsim \sqrt{\frac{\alpha_b }{R_b\gamma_b}}\gamma_e^{3/2}. \label{eq:stab_cond}
\end{equation}
Note that this expression is equivalent to $\Psi_{Fb} \ll \omega_{pb}$ translating into the decoherence of the beam, as invoked in LP11. According to B04 two-population analysis, including electron temperature further inhibits the growth of $\mathrm{F_1}$ modes:
\begin{equation}
    \rho_b > \sqrt{\frac{\alpha_b\gamma_e^3}{R_b\gamma_b}}\frac{\rho_{e\perp}}{\sqrt{\rho_{e\perp}^2-\rho_{e\parallel}^2/3}}\sqrt{1-(\rho_{e\perp}^2-\rho_{e\parallel}^2/3)}, \label{eq:rhobiso}
\end{equation}
The isotropic electron-temperature parameter $\rho_e \equiv u_{th}/u_b$ can be shown to lower the beam-temperature stabilization threshold.
This result is the opposite of what was found by the authors of LP11, who found an enhancement of $\mathrm{F_1}$ modes with increased thermal spread of an ultra-relativistic Maxwellian distribution. The discrepancy is not to be attributed to the different distribution functions chosen but instead, to the finite drift velocity of electrons considered in B04, and not in the model of LP11. While different thermal effects are not expected when applying what appears to be a simple transformation to the electron rest frame, the finite drift of background ions can not be neglected anymore. In other words, electrons may act as the driving population as the beam loses coherence by dispersion effects, until they are themselves thermalized and all instabilities are quenched. In the model of LP11, the driving population is always the beam, whose plasma frequency consistently determines the wavenumber cut-off, but at the expense of the kinetic quenching predicted in B04. That is because electron temperature would not be as important and may in fact enhance the FI by making the electrons heavier, thus relaxing the stabilization condition \eqref{eq:rhobiso} by effectively decreasing the mass-ratio $R_b$. This effect will be recovered with a three-population model that can as well solve the discrepancies discussed here.

\subsubsection{Solution: three-population model}
\begin{figure}
	\includegraphics[width=\columnwidth]{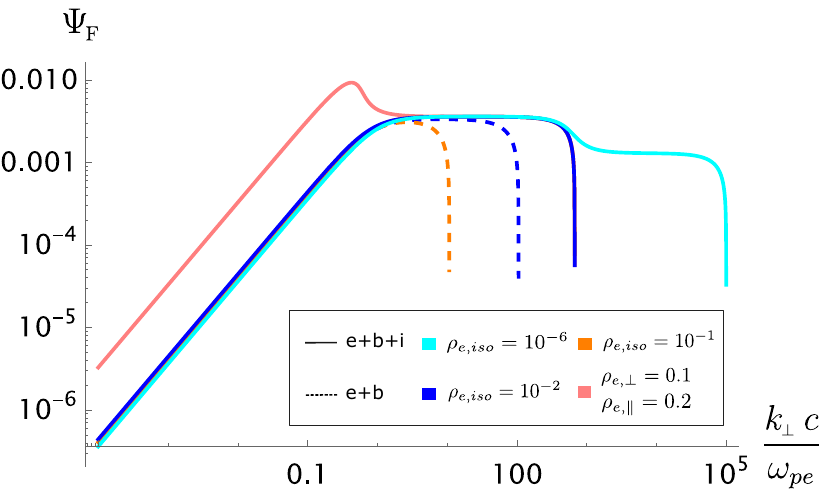}
    \caption{One-dimensional spectrum of the filamentation instability of $k_\perp$ perturbations at fixed $\rho_b=10^{-5}$, $R=1836$, $\alpha_b=0.1$. Dashed resp. solid lines are solutions to the two-population ``e+b'' resp. three-population ``e+b+i'' dispersion relations. Colors correspond to different isotropic electron temperatures, except for the pink line. The latter represents a situation $\rho_{e\parallel}>\rho_{e\perp}$ where anisotropy amplifies the growth rate of $\mathrm{F_1}$ modes at the beam inertial scale $k_\perp=\omega_{pb}/c \simeq 0.45\omega_{pe}/c$ for the chosen parameters.}
    \label{fig:Spectrum_FI}
\end{figure}

With applications to the precursor of a relativistic shock in mind, a cold background ion population is added to the dispersion relation in its own inertial frame. The dielectric component $\varepsilon_{xz}$ remains unaffected by a static population while diagonal terms $\varepsilon_{xx}$ and $\varepsilon_{zz}$ are appended by a single term $-R_i$. The threshold wavenumber $k_c$ in a cold plasma (25) is modified as \begin{equation}
    \frac{k_c c}{\omega_{pb}}=\frac{1}{\rho_b\sqrt{\gamma_b}}\qty(1-\frac{\gamma_b\rho_b^2 R_b}{\alpha_b \gamma_e^3}\qty(1+\frac{\gamma_e^3}{R_i}\qty(1+\frac{\alpha_b}{\gamma_b^4})-\frac{\alpha_b^2 \gamma_e^2 - \frac{\rho_{p\parallel}^2}{3}}{\rho_{p\perp}^2}))^{1/2}
\end{equation}
Note first that  parallel electron temperature can be neglected for $\rho_{p\parallel}\ll 3 \alpha_b^2\gamma_e^2$ i.e $u_{th,e}\ll u_e$. Then, unlike the two-population model, the last unstable wavenumber is found beyond the FI decoherence cut-off determined by $\Psi_{\mathrm{F_1}} \ll \omega_{pb}$, indicating the transition to small-scale $\mathrm{F_2}$ modes arising from the interaction within the background plasma (see plateau at large $k_\perp$ for the cyan line in \autoref{fig:Spectrum_FI}). \\ New kinetic effects are also unveiled. Increasing plasma temperature isotropically stabilizes the smaller-scale $\mathrm{F_2}$ modes thus adding a second cut-off after the $\mathrm{F_1}$ one. When increasing $\rho_{p\parallel}$ at fixed $\rho_{p\perp}$, the unstable domain is increased. A $\rho_{p\parallel}$-dominating anisotropy is also associated with an enhancement of the maximal growth rate of the $\mathrm{F_1}$-spectrum at the beam inertial scale $k_x \sim \omega_{pb}/c$ (see solid blue line on \autoref{fig:Spectrum_FI}). 

We can therefore reconcile the results of LP11 and B04 by taking into account the background ion population in the upstream frame, at the same time unveiling the existence of secondary small scale instabilities whose role can not be overlooked in the kinetic regime as well as in the high beam density $\alpha_b \rightarrow 1$ cold regime.

\section{Application to transrelativistic shocks}
\label{sec:transre}
In the previous sections, the general linear theory of electrostatic and electromagnetic instabilities triggered by a relativistic beam have been presented. While electron temperature parameters relate to the inner dynamics of the plasma, the beam Lorentz factor $\gamma_b$ and density $\alpha_b$ are parameters that must be given from the configuration of the shock. For this purpose, a model of the beam based on isotropic reflection in the shock-frame is presented in Appendix \ref{sec:A_K}
\\~\\
The energy gained by particles reflected at the shock depends on the reflection process through the cosine of the angles at which the particle crosses the shock $\mu_{\rightarrow sh|u}$ and $\mu_{u\leftarrow |sh}$ via the transformation \citep{Gallant1999}: 
\begin{equation}
    \gamma_{b|u}=\Gamma_{sh}^2 (1+\beta_{sh}\mu_{\rightarrow sh|u})(1+\beta_{sh}\mu_{u\leftarrow |sh}),
    \label{eq:E_beam}
\end{equation}
The maximum energy is then attained by particles that travel up to a gyroradius \citep{Lemoine2011}\begin{equation}
    r_{L|u}^{(max)}\simeq \frac{4\Gamma_{sh}^2m_i c^2}{ZeB_u}
\end{equation} 
Eventually, particles traveling along the Larmor's circle are caught up by the shock traveling at $\beta_{sh}$ in the upstream frame. A shock-crossing time scale can thus be derived as a lower limit for the growth rate of instabilities in the precursor. While it depends on the phase velocity of a given wave, an estimate for perpendicular-propagating modes is \begin{equation}
    \frac{\tau_{c|u}^{-1}}{\omega_{pe}}\simeq 2\times 10^{-7} \beta_{sh}\Gamma_{sh}Z\qty(\frac{R_b}{1836})\qty(\frac{n_e}{\textrm{cm}^{-3}})^{-1/2}\qty(\frac{B_u}{\mu G}). \label{eq:shock_cross}
\end{equation}
It appears as a prohibitive red line, limiting the growth of the shock-perpendicular modes (e.g filamentation modes or buneman oblique in the cold limit) on \autoref{fig:Transrelat}. 
\subsection{Limited Fluid growth in the precursor}

\subsubsection{Beam energy $\gamma_b(\Gamma_{sh})$}
In the two previous sections were obtained the growth rates for electrostatic and electromagnetic instabilities, respectively. In the cold approximation, primary modes were shown to depend explicitely on the beam Lorentz factor as $\Psi_{\mathrm{B_1}\parallel} \propto \gamma_{b}^{-1}$ and $\Psi_{\mathrm{B_1}\perp} \propto \gamma_{b}^{-1/3}$ for the two limit electrostatic buneman modes, and $\Psi_{\mathrm{F_1}} \propto \beta_{b}\gamma_b^{-1/2}$ for the perpendicular filamentation mode. Weaker dependence of the perpendicular modes is a direct consequence of boosted dielectric response in  perpendicular directions. The mean beam velocity is derived as a function of the shock's Lorentz factor $\Gamma_{sh}$ from the model's exact expression \eqref{eq:beam4veloc}:
\begin{align}
    \gamma_b(\Gamma_{sh}) \simeq \begin{dcases} 3\Gamma_{sh}^2, &\Gamma_{sh} \gg 1 \\ 1+\frac{3}{2}\beta_{sh},& \beta_{sh} \ll 1 \end{dcases} 
\end{align}
\subsubsection{Density-ratio transformation $\alpha_b(\Gamma_{sh})$}
As primary instabilities arise from the interaction between electrons and beam ions, growth rates are a direct function of $\alpha_b$, which also depends on $\Gamma_{sh}$. Indeed, upstream ions being isotropized by elastic collisions in the shock-transition layer, the reflection rate $\alpha_{i|sh}$ must be defined a constant model parameter in the shock-frame. It is then transformed to the upstream frame,
 \begin{equation}
    \alpha_{i|u}=\Gamma_{sh}^2\alpha_{i|sh}, \label{eq:ai_transfo}
\end{equation}
and the beam-to-electron density ratio $\alpha_b \equiv \alpha_{e|u}$ is obtained using total charge neutrality (Z=1) $ n_e=n_b+n_i$:
 \begin{equation}
    \alpha_b=\dfrac{\alpha_{i|u}}{1+\alpha_{i|u}}. \label{eq:def_aeai}
\end{equation}
This framework avoids superluminal electron response as $\alpha_b \leq 1$, converging to unity in the ultrarelativistic range regardless of the value of $\alpha_{i|u}$. This new result is reflected on the scaling of the primary Buneman mode $\Psi_{\mathrm{B_1}} \propto \alpha_b^{1/3}$, and on the transition of dominant electromagnetic mode from the primary $\mathrm{F_1}$ to the secondary $\mathrm{F_2}$ seen on \autoref{fig:Transrelat}. 

\subsubsection{Analysis of the growth rates $\Psi(\Gamma_{sh})$}
\begin{figure}
	\includegraphics[width=\columnwidth]{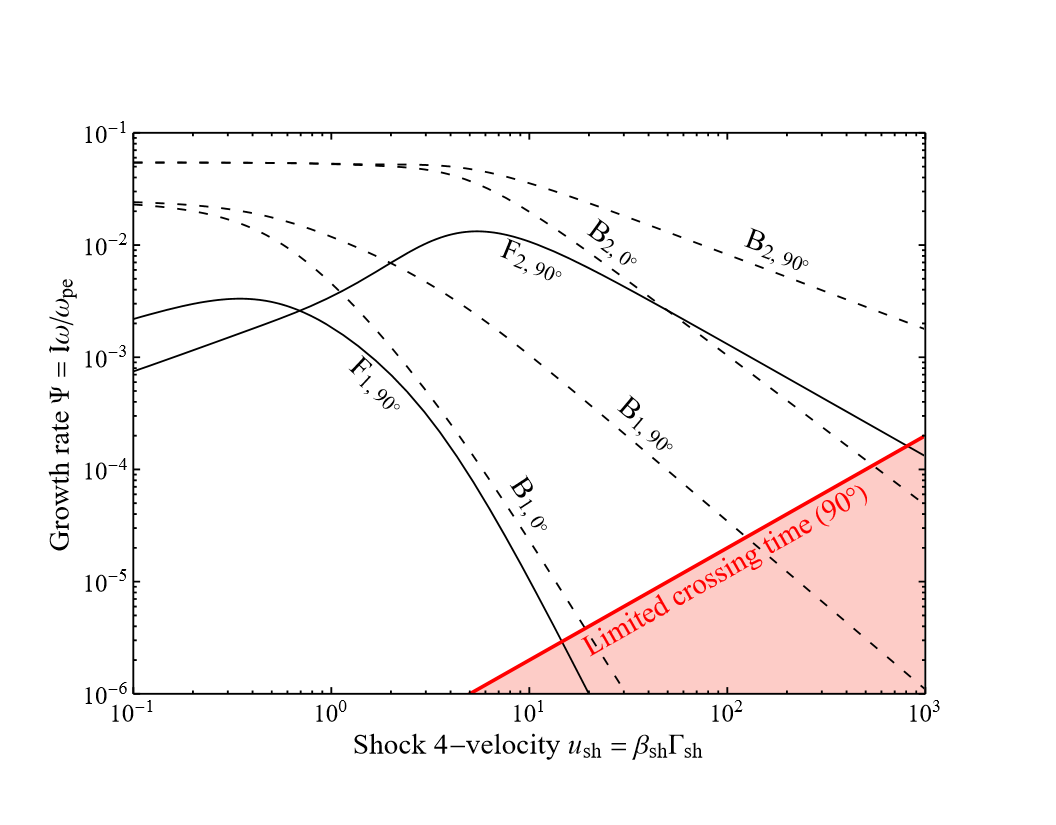}
    \caption{Growth rate for the Buneman (dashed lines) and filamentation (solid lines) modes in the fluid limit, over the transrelativistic range. The subindex of a curve's label indicates the primary of secondary nature of the instability as well as the wave-vector angle in degrees. Chosen parameters are $\alpha_{i|sh}=0.1$, $R_0=1836$ and no velocity dispersion in the beam nor temperature of electrons are considered in this limit. The exact analytical expressions \eqref{eq:BI_growth} and \eqref{eq:growth_FI_cold} are used, and adapted for the secondary instabilities. The red region defines the region where the growth of perpendicular-propagating modes is limited by the precursor-crossing time \eqref{eq:shock_cross} with typical ISM parameters.}
    \label{fig:Transrelat}
\end{figure}

In the fluid limit, where the beam and plasma are treated as cold interpenetrating beams, the growth rate of electromagnetic instabilities depends on $\Gamma_{sh}$ via the beam energy $\gamma_b$, and beam-to-electron mass and density ratios, $\alpha_b$ and $R_b$ respectively. In the ultra-relativistic limit, the asymptotic scalings are given only by $\gamma_{\{b,e\}}^{-1/2}$ for $F_{\{1,2\}}$ modes, $\gamma_{\{b,e\}}^{-1}$--- $\gamma_{\{b,e\}}^{-1/3}$, for the oblique continuum of $B_{\{1,2\}}$ modes. The beam's lorentz factor was given above as $\gamma_b \sim \Gamma_{sh}^2$ and $\gamma_e \simeq \alpha_{i|sh}\gamma_b^{1/2}\sim \Gamma_{sh}$ for drifting electrons by current neutrality in the $\alpha_b \rightarrow 1$ limit.\\

Consistently with what is seen on \autoref{fig:Transrelat}, electron heating is expected to occur in electrostatic Buneman waves. Within the fluid approximation employed here however, growth rate for $\mathrm{B_1}$ remains always larger than for $\mathrm{F_1}$  throughout the whole transrelativistic range and no transition from dominant primary instability to another is observed in the \textit{initial} (i.e cold) parameter space. This new result is not trivial for one would expect a transition if inferring only based on beam Lorentz factor $\gamma_b$ dependence. The curves of $\mathrm{B_1}$ and $\mathrm{F_1}$ not crossing is a consequence of the self-consistent $\alpha_b$ transformations introduced here. We therefore argue that the transition would happen instead \textit{dynamically}, piloted by the influence of self-generated kinetic effects on the growth of electromagnetic instabilities.

\subsection{Kinetic quenching in a cold precursor}
Kinetic effects can indeed modify the picture presented on \autoref{fig:Transrelat} in quite subtle aspects. Not only is a relativistic dispersion relation with finite temperatures hardly tractable for analytical treatment, it would not make sense to solve it numerically as quantitative results are limited by the linear approximation. Instead, key insights are readily available by substituting $\rho_b(\Gamma_{sh})$ in the constraints derived in \autoref{sec:FI_kin}.
\\~\\
Perpendicular velocity dispersion in the beam, as determined from the reflection model,
\begin{equation}
    \rho_{b\perp}\equiv \frac{\Delta u_\perp}{u_b}\simeq\begin{dcases}\frac{2}{3}\frac{1}{\Gamma_{sh}},&\Gamma_{sh}\gg 1 \\
    \sqrt{2},&\beta_{sh}\ll 1   
    \end{dcases}
\end{equation}
has been shown to quench the FI according to the stabilization condition \eqref{eq:stab_cond}, which can now be reduced to the simple, Lorentz-invariant form 
\begin{equation}
   \sqrt{\frac{\alpha_b}{R_b}} \leq \frac{2}{3}~~~(\Gamma_{sh} \gg 1). \label{eq:stab_cond_3}
\end{equation}
It has to be noted that $\alpha_b$ is kept unconstrained here because the $\Gamma_{sh}$ scaling of $\rho_{b\perp}$ cancels with $\gamma_b$ much faster than $\alpha_b$ would tend to unity for typical values of $\alpha_{i|sh}$. \\~\\
Now going further in the relativistic regime, $\Gamma_{sh}\gg \alpha_{i|sh}^{-1/2}$ (i.e $\alpha_b \simeq 1$), one can reduce the \eqref{eq:stab_cond_3} in the upstream frame to essentially $\alpha_{sh}^2\Gamma_{sh}^2 \leq R_0$. Considering a hydrogen plasma $R_0=1836$, any $\alpha_{sh}\leq 1$ gives a lower bound $\Gamma_{sh}^{-} \simeq 100$ (consistently with what can be found in the literature, with varying approaches \citep{Lyubarsky2006,Rabinak2011, Lemoine2011} which is greater than the upper bound $\Gamma_{sh}^{+}\simeq 15$ set by shock-crossing time. Therefore, a cold precursor with baryonic load equal to or equivalent to a hydrogen plasma can not admit large-scale magnetostatic turbulence\footnote{Secondary, smaller-scale $\mathrm{F_2}$ modes may still appear up to $\Gamma_{sh}\sim 1000$}. For a pair-plasma $R_0=1$ however, $\mathrm{F_1}$ modes are allowed for $\Gamma_{sh} \overset{\sim}\in [\alpha_{sh}^{-1},50]$, which is a non-null interval for realistic reflection ratios $\alpha_{sh} \geq 2\times 10^-2$ . 
\\~\\
Although the stabilization condition considered together with the shock-cross time isn't completely prohibitive for a pair-plasma, clearly a larger range of $\Gamma_{sh}$ are observed to be capable of generating large-scale magnetostatic turbulence for various mass ratios in PIC simulations \citep{Sironi2011,Vanthieghem2022}.  This prompts further consideration of electron temperature effects of electron temperature on the growth rate of $\mathrm{F_1}$ modes.

\subsection{Discussion on the role of electron preheating}
\label{subsec:preheating}
So called \textit{anomalous} electron heating may indeed play a role, not only to account for the observed energy of gamma-ray emission, but also as a necessary condition to trigger the growth of a persistent magnetic structures in the precursor of transrelativistic shocks. It should naturally lead the reader to contemplate the potential role of electrostatic instabilities in pre-heating the precursor to temperature conditions favourable for the growth of large scale magnetostatic turbulence.
\\~\\
In support of this conjecture, it is suprising how many insights fetched along this analysis point synergistically towards a key role of electrostatic modes in facilitating the later growth of electromagnetic modes. First, the Buneman instability, in the absence of early non-linear saturation, may heat up the plasma up to $\gamma_{th,e} m_e \simeq \gamma_b m_b$ so as to make any plasma in the precursor effectively behave as a pair-plasma $R_b \simeq 1$ and therefore an efficient generator of persistent magnetostatic modes according to \eqref{eq:stab_cond_3}. Second, electron temperature was shown to relax the kinetic quenching in \eqref{eq:rhobiso}, thus facilitating the growth of large-scale filamentation in the non-relativistic regime. Third, \autoref{fig:BIandBII_3} shows the obliquity of dominant modes progressively converging towards beam-parallel directions with increasing temperature. This would dynamically induce a slightly stronger parallel electron temperature, that was shown in \autoref{fig:Spectrum_FI} to significantly enhance the growth of large-scale $\mathrm{F_1}$ modes at the beam skin depth $k_\perp=\omega_{pb}/c$. \\~\\
Note also that the growth of electrostatic and secondary electromagnetic instabilities remains unaffected by the constraints set by the analysis. The role of small-scale, fast-growing secondary instabilities on the early dynamics of the plasma in shaping the precursor of relativistic shocks must be further investigated.

\section{Summary and discussion of the results}
\label{sec:summary}
Here, main results from this study are summarized.
\begin{itemize}
    \item \textit{Primary Buneman instability}: a generalized derivation procedure has been applied to obtain the growth rate of relativistic two-stream like instabilities in a cold \eqref{eq:BI_growth} and hot \eqref{eq:growth_BI_hot} plasma, for arbitrary angles of propagation and mass ratios. Including electron temperature in the dispersion relation, a relativistic version of the Penrose stability criterion, $u_e + u_{th,e} \geq u_b$ \eqref{eq:Penrose_BI}, is found. Oblique modes are also shown to dominate over parallel modes due to electric field boost in directions transverse to the beam-axis. \\
    \item \textit{Secondary ion-electron instabilities}: including the interaction of background ions in the dispersion relation, previously overlooked, unveils the existence of secondary Bunemann and Filamentation instabilities understood to arise from the interaction of drifting electrons with static ions within the background plasma. The secondary, small-scale, Buneman modes always grows faster than the primary modes albeit stabilized at low temperatures $u_{th,e} \geq u_e$ \eqref{eq:criterion_BII}. Similarly, the secondary filamentation modes takes over primary $\mathrm{B_1}$ and $\mathrm{F_1}$ modes due to a comparatively smaller velocity difference induced by electrons drifting in the background plasma. \\
    \item \textit{Filamentation instability}: an updated analysis of the filamentation instability with a three-population dispersion relation resolves inconsistencies in the analytical treatment of kinetic effects in the high-beam density limit \eqref{eq:F_treshc}. New solution for the continuous spectrum of $\mathrm{F_1}$ and $\mathrm{F_2}$ modes is calculated (\autoref{fig:Spectrum_FI}), and a targeted enhancement of the $\mathrm{F_1}$ maximum growth rate is observed at large scales $k_\perp \simeq \omega_{pb}/c$ when electron temperature is parallel-anisotropic.\\
    \item \textit{Dominant instability mediating relativistic shocks}: a hierarchy of beam-plasma instabilities is obtained over the transrelativistic range by constraining equilibrium distribution parameters with a relativistic model for ion reflection at the shock (\autoref{fig:Transrelat}). Electrostatic Buneman modes are shown to dominate in non-relativistic shocks, while secondary filamentation and oblique Buneman modes take over as soon as for mildly relativistic shocks. No direct transition from $\mathrm{B_1}$ to $\mathrm{F_1}$ as the dominant large-scale mode is observed in the fluid plasma approximation. \\
    
    \item \textit{Magnetostatic turbulence over the transrelativistic range}: kinetic quenching of $\mathrm{F_1}$ modes due to perpendicular beam-spread \citep{Lemoine2011} is obtained \eqref{eq:rhobiso} and used for an updated calculation of the minimal Lorentz factor $\Gamma_{sh}^{(-)} \simeq \alpha_{i|sh}^{-1}\sqrt{R_0}$ admitting the growth of $\mathrm{F_1}$ modes. Further stringent constraints are obtained including limited precursor-crossing time, with ultra-relativistic shocks $\Gamma_{sh}>\Gamma_{sh}^{(+)}\simeq 15$---$50$ shown to prohibit large-scale magnetostatic turbulence growth in cold, baryon-loaded ($R_0 \geq 1836$) precursors.\\
\end{itemize}
The results here itemized are to be considered within their range of application. Before anything else, the framework of linear theory prevents a quantitative investigation of the role of electrostatic instabilities in electron pre-heating, and conjecturing so (\autoref{subsec:preheating}), in piloting the transition to secular electromagnetic filamentation growth. Therefore, electron temperature is limited to the derivation of stability criteria for electrostatic instabilities and qualitative analysis of kinetic effects on the growth of electromagnetic modes. To avoid non-analytical and resp. non-alegbraic dispersion relations, the waterbag resp. low temperature approximation $\rho_e \ll 1$ have been used. They should in fact constitute a single approximation within a linear theory's range of applicability (see also the conclusion of \cite{Bret2005} for quantitative discussion on the matter). \\~\\
A waterbag function was also used to represent perpendicular velocity dispersion in the beam, though it happens to be an exact description of the truncated uniform distribution obtained from isotropic reflection in the shock-frame. The dispersion relation for filamentation modes was written in the low-temperature limit as-well, but then again, the limit for this approximation lies beyond the stabilization criterion of all modes \eqref{eq:stab_cond}. Isotropic reflection in the shock frame is also one assumption used for application to transrelativistic shocks (\autoref{sec:transre}) that could easily be challenged and adapted to upgrade the model presented in \autoref{sec:A_K}. Considering a cold initial crossing, and overlooking the feedback of self-generated turbulence on the shock reflection process are other assumptions however tied to the consideration of linear response.

\section{Conclusion}
\label{sec:conclusion}
This study presents a preliminary step in understanding the stability of relativistic shock precursors. It offers a useful reference point for future research aiming to extend the transrelativistic model to various scenarios, including different background magnetic fields, non-uniform density distributions, deviations from charge or current neutrality, and anisotropic reflections. While the detailed role of each instability in the dynamics of the plasma remains out of reach of an analytical treatment,  an important finding emerges: the initial conditions of relativistic shocks may not fully account for the observed electromagnetic turbulence in kinetic simulations, crucial for the onset of a \textit{Fermi} acceleration process. Therefore, considering kinetic effects associated with electron preheating in self-generated turbulence is a minimal requirement to account for the further large-scale development of the filamentation instability. Accordingly, a scenario where  the persistence of turbulent electric fields parallel to the shock could sustain the growth of large-scale electromagnetic modes was proposed. As far as numerical simulations are concerned, this is only a matter of interpreting the results. 
\\~\\
The existence of rapidly growing, small-scale secondary modes however raises concerns about the accuracy of current findings. Secondary Buneman modes might contribute to preheating but could evade detection by conventional numerical schemes that filter out scales smaller than the plasma skin-depth \citep{Godfrey2015}. This practice, aimed at studying the precursor's long-term dynamics while mitigating numerical instabilities at high wave numbers, may inadvertently lead to incomplete representations of plasma dynamics. Consequently, it prompts inquiries into whether kinetic simulations of relativistic shocks underestimate electron heating at small scales, and thereby inaccurately reflecting the turbulent content of the precursor. 
\\~\\
To address this query, further numerical investigations focusing on the initial plasma dynamics at small scales are needed, particularly within the relativistic framework. Existing high-resolution simulations (e.g \citet{Keshet2009,Vanthieghem2022}), could already shed light on the presence of secondary electrostatic modes. These modes, akin to parallel modes but expected to persist longer, exhibit a unique behavior—they are roughly carried along with the flow in mildly relativistic shocks, setting them apart from primary Buneman waves that move with the beam. Additionally, oblique variants of primary Buneman modes demonstrate a similar growth rate enhancement, suggesting their potential role in perpendicular heating near the shock-transition layer.
\begin{acknowledgments}
This work was partially supported by Fundação para a Ciência e a Tecnologia (FCT-Portugal) through Contract No. UI/BD/154835/2022.
\end{acknowledgments}

\appendix

\section{Two-stream-like instability}
\label{sec:A_TS}
This appendix details the derivation for the spectrum of unstable modes susceptible to a two-stream-like instability between to cold, drifting species with arbitrary mass and density ratios, $R_b$ and $\alpha_b$. The original procedure due to \cite{Bludman1960} is extended to relativistic drift velocities, and reported in B04. Here, the calculation are updated to include waterbag kinetic effects within an electrostatic framework. The following approach could be adapted to waves with finite $k_x$, but will be constrained to one-dimensional langmuir waves propagating along the beam axis to keep the analysis tractable.

\subsection{Dispersion relation}
Consider the 1D susceptibility for thermal electrons under the waterbag approximation \eqref{eq:bunneman_hot1D}. It can be expanded to first order in y=$\omega-k_zV_b$, and included into the dispersion relation \eqref{eq:bunemanfull_2D} to find \begin{equation}
    \omega_{pe}^{-2} - \qty(2k_z^2c^2u_{th} )^{-1}\qty(\frac{1-\frac{y}{k_zc\beta_+}}{\beta_b-\beta_+}-\frac{1-\frac{y}{k_zc\beta_-}}{\beta_b-\beta_-})-\frac{\alpha_b/R_b}{\gamma_b^3}\frac{1}{y^2}.
\end{equation}
Recall that the lower and upper thermal velocities $\beta_\pm$ are given by \eqref{eq:bpm}. The dispersion relation can be re-arranged into a polynomial equation of the form:
\begin{equation}
    Y^3/(K_\parallel C_3)^3 + Y^2\qty(1-(K_\parallel C_2)^{-2})-\Psi_0^3=0, \label{eq:A_disp}
\end{equation}
with \begin{align}
    &K_\parallel\equiv \frac{k_z c}{\omega_{pe}};\\
    &C_2^2=\frac{\qty(\beta_b - \beta_+)^{-1}-\qty(\beta_b-\beta_-)^{-1}}{2u_{th}};\\
    &C_3^3=\frac{\qty(\beta_b - \beta_+)^{-2}-\qty(\beta_b-\beta_-)^{-2}}{2u_{th}};\\
    &\Psi_0^3=\frac{\alpha_b/R_b}{\gamma_b^3}.
\end{align}
The dispersion relation \eqref{eq:A_disp} is a depressed cubic equation, which can be solved following the aformentioned references. First assume a complex solution of the form \begin{equation}
    Y=r\exp{i\phi},
\end{equation}
taking the real part of \eqref{eq:A_disp} and injecting it into the imaginary part yields \begin{align}
    &2^{1/3}\qty(\frac{C_2}{C_3})^2\qty(1-(K_zC_2)^2)\Psi_0^{-2}=\mathcal{C}(\phi), \label{eq:A_Ceq}\\
    &\mathcal{C}(\phi)\equiv \frac{\sin 3\phi}{\sin 2\phi}(-\cos\phi)^{1/3}.  \label{eq:A_Cdef}
\end{align}
This equation gives the expression of the argument $\phi$ as a function of $\omega$, $k_z$ and model parameters.\\

We recover a result equivalent to that of \cite{Bret2004} for cold electrons in the non-relativistic limit: \begin{equation}
-\frac{\qty(\tilde{Z}_z^2\gamma_e^3-1)}{2\gamma_e \Psi_0^2}=\mathcal{C}(\phi),
\end{equation}
introducing the adimensional parameter $\tilde{Z_z}\equiv k_zV_b/\omega_{pe}$.

\subsection{Solution and stabilization}
The dispersion relation is eventually solved as \begin{align}
    &Y=\frac{K_z C_3}{2^{1/3}}\Psi_0^2\mathcal{G}(\phi),
    &\mathcal{G}(\phi) \equiv (-\cos\phi)^{1/3}\exp{i\phi(K_z)}. \label{eq:A_gTS}
\end{align}
While the argument can now be expressed exactly as a function of $K_z$ by inverting \eqref{eq:A_Cdef}, it is enough to note that $\mathcal{G}(\phi)$ admits a periodic maximum in $\phi_m=-2\pi/3 + 2\pi n,~n\in\mathbb{Z}$ with $G(\phi_m)=\frac{\sqrt{3}}{2^{4/3}}$.\\~\\
The temperature correction to the cold growth rate is thus completely encoded in the quantity $C_3(u_{th}) \xrightarrow[u_{th} \to 0]{}2^{1/3}$. It is immediately verified that $\mathcal{C}(\phi_m)=0$, such that the solution to \eqref{eq:A_Ceq} gives the maximum-growing wavenumber 
\begin{equation}
    K_{z,m}=1/C_2(u_{th}). \label{eq:A_thresh}
\end{equation} 
In the cold limit, a value close to the cerenkov resonance is confirmed \begin{equation}
   \lim_{u_{th} \to 0} K_{z,m}V_b= \gamma_e^{-3/2}(1-\alpha_b)^{-1} \overset{\alpha_b\ll 1}{\simeq} 1.
\end{equation}
\section{Filamentation instability}
\label{sec:A_FI}
The dispersion relation of \cite{Bret2004} updated to a three-population model (only adding a $-1/R_i$ contribution in $\varepsilon_{zz}$ and $\varepsilon_{xz}$ terms) is written with normalized quantities:
\begin{equation}
\hspace{-2cm} 
\begin{split}
\left(-\frac{\alpha_b}{\gamma_b R_b \left(\Omega^2-\rho_{b\perp}^2 Z_\perp^2\right)}-\frac{1}{\gamma_p \left(\Omega^2-\rho_{e\perp}^2 Z_\perp^2\right)}-\frac{1}{R_i
   \Omega^2}+1\right) & \left(-\frac{Z_\perp^2\alpha_b^2 \gamma_p^2 \left(1+\frac{\rho_{e\parallel}^2}{3}-\rho_{e\perp}^2\right)+\Omega^2}{\gamma_p^3
   \left(\Omega^2-\rho_{e\perp}^2 Z_\perp^2\right)}-\frac{\alpha_b Z_\perp^2}{\gamma_b R_b \left(\Omega^2-\rho_{b\perp}^2 Z_\perp^2\right)}-\frac{\alpha_b}{\gamma_b^3
   R_b}-\frac{Z_\perp^2}{\beta^2}-\frac{1}{R_i}+\Omega^2\right)\\& -\alpha_b^2 Z_\perp^2 \left(\frac{1}{\gamma_b R_b \left(\Omega^2-\rho_{b\perp}^2
   Z_\perp^2\right)}+\frac{1}{\gamma_p \left(\Omega^2-\rho_{e\perp}^2 Z_\perp^2\right)}\right)^2=0 \label{eq:FI_kin}
\end{split}    
\end{equation}
Frequency and perpendicular wave-numbers are normalized as $\Omega\equiv \frac{\omega}{\omega_{pb}}$ and $Z_\perp \equiv k_x V_b/\omega_{pb}$, while a temperature parameter $\rho_{s\mu}\equiv \frac{\Delta_\mu u_s }{u_s}$ is introduced for the beam and electron species. For non-relativistic responses $\rho_{s\mu} \ll 1$, the $\parallel$ component would only appear in the form of $(1+\rho_{s\parallel}^2/3-\rho_{s\perp}^2$). The reflection model in Appendix \ref{sec:A_K} prescribes beam-parallel temperatures well within $\rho_{b\parallel}^2/3 \ll 1$, such that it can be completely neglected in the dispersion relation. It also means for the electron response in \eqref{eq:FI_kin} to be considered quantitatively only for non-relativistic temperature $\rho_{e\mu} \ll 1$. This range of validity is methodically respected in the analysis of \autoref{sec:FI_kin}.

\section{Relativistic kinematics of the shock}
\label{sec:A_K}

\subsection{Relativistic beaming of returning ions}
The model presented here is not concerned with an estimate of the fraction of ions reflected at the shock, which is encoded in the parameter $\alpha_b$, but rather to constrain beam distribution parameters $\gamma_b$ and $\rho_b$ as a function of the shock's Lorentz factor $\Gamma_{sh}$. As these initial parameters characterize the beam's distribution at  equilibrium, minimal assumptions are chosen within a linear's theory range of applicability. That is, the feedback of instabilities driven by the beam on itself is neglected in the model and the reflection process is in its very first instants. The first implies that the distribution obtained for the beam is indeed a constant equilibrium distribution. The latter allows to consider the reflection process starting with low entropy ions, i.e starting with $\mu_{\rightarrow sh}=1$ in \eqref{eq:E_beam}:
\begin{equation}
\gamma_b(\mu_{sh})=\Gamma_{sh}^2(1+\beta_{sh})(1+\beta_{sh}\mu_{sh})
\end{equation}
$\mu_{sh}=$
are isotropized in the shock-transition layer. In turn, this implies a uniform distribution of returning angles $\mu_{u\leftarrow sh|}\in[0,1]$. Performing a Lorentz transformation to the upstream frame, the half-sphere of returning ions is further truncated according to relativistic aberration \citep{Rybicki1985},
\begin{equation}
    u_\perp=\mathcal{A}(\mu_{u\leftarrow |sh})u_{\parallel}, \label{eq:aberration}
\end{equation}
where \[\mathcal{A}(\mu_{u\leftarrow |sh})\equiv\frac{1}{\Gamma_{sh}}\frac{\sqrt{1-\mu_{u\leftarrow |sh}^2}}{\mu_{u\leftarrow |sh}+\frac{\beta_{sh}}{|\beta_{b|sh}|}}\]
is the tangent of the returning angle in the upstream frame. The quantity $|\beta_{b|sh}|$ is the norm of the beam velocity in the shock frame, simply given by \begin{equation}
    \beta_{b|sh}\qty(\Gamma_{sh})=\sqrt{1-\dfrac{1}{\left(1+\beta_{sh}\right)^2 \Gamma_{sh}^2}}.
\end{equation} Naturally, the perpendicular component is unchanged along the beam axis, $\mathcal{A}(\mu_{u\leftarrow |sh}=1)=0$. The ions travelling along the shock surface in its comoving frame are beamed within a cone of opening 
\begin{equation}
    \mathcal{A}(\mu_{u\leftarrow |sh}=0)=\frac{\beta_{sh}}{|\beta_{b|sh}|}\frac{1}{\Gamma_{sh}}.
\end{equation}

The factor $\beta_{sh}/|\beta_{b|sh}|$ generalizes the ultra-relativistic scaling $u_\perp \simeq u_\parallel/\Gamma_{sh}$ to transrelativistic shocks.  

The parallel-component of the four-velocity also admits dispersion via the norm's and aberration dependence on $\mu_{sh}$:
\begin{equation}
    u_\parallel(\mu_{sh})= \qty(\frac{\gamma_b(\mu_{sh})^2-1}{1+\mathcal{A}(\mu_{sh})^2})^{1/2}.
\end{equation}
As the reflection process is a truncated uniform distribution, the distribution of velocities of particiles in the beam will also be distributed according to a waterbag distribution. As expected from the beaming effect, energy is focused along the shock normal axis such that in the ultra-relativistic limit $\Gamma_{sh}\rightarrow \infty$,  $\langle u_\perp \rangle \rightarrow 0$ and $\langle u_\parallel \rangle \rightarrow \langle u_b \rangle$. A new, non trivial result is obtained for the beam-parallel dispersion,
\begin{equation}
    \rho_{b\parallel}\equiv \frac{\Delta u_\parallel}{\overline{u_\parallel}}\simeq\begin{dcases}\frac{1}{3},~~\Gamma_{sh}\gg 1 \\
    1,~~\beta_{sh}\ll 1   \label{eq:rhob}
    \end{dcases}
\end{equation}
smoothly decreasing from unity to $\sim 1/3$ in ultra-relativistic limit. This result is key to reduce the model to verifying consistently $\rho_b \ll 3$, such that parallel dispersion can be completely neglected in the dispersion of the filamentation instability. The quantities that are used for the kinetic model of the beam are therefore the mean parallell velocity 
    \begin{equation}
    u_b \equiv \overline{u_\parallel}\simeq\begin{dcases}
        3\Gamma_{sh}^2,~~\Gamma_{sh}\gg 1 \\
        \sqrt{\beta_{sh}},~~~~~~~~~~~~\beta_{sh}\ll 1
    \end{dcases}
\end{equation}
and the perpendicular dispersion
 \begin{equation}
    \rho_{b\perp}\equiv \frac{\Delta u_\perp}{u_b}\simeq\begin{dcases}\frac{2}{3}\frac{1}{\Gamma_{sh}},~~\Gamma_{sh}\gg 1 \\
    \sqrt{2},~~\beta_{sh}\ll 1   
    \end{dcases}
\end{equation}
Exact transrelativistic expressions are used for plots, calculated from: 
\begin{equation}
    \Delta u_\perp\simeq \dfrac{\beta_{sh} \sqrt{\dfrac{\left(\beta_{sh}+1\right)^2 \Gamma_{sh}^4-1}{\dfrac{1-\dfrac{1}{\Gamma_{sh}^2}}{1-\dfrac{1}{\left(\beta_{sh}+1\right)^2 \Gamma_{sh}^2}}+1}}}{\sqrt{1-\dfrac{1}{\left(\beta_{sh}+1\right)^2 \Gamma_{sh}^2}}}
\end{equation}
and 
\begin{equation}
    \begin{matrix}u_b\qty(\Gamma_{sh}) \\ \Delta u_\parallel\qty(\Gamma_{sh}) \end{matrix} =\dfrac{1}{2} \left(\sqrt{\left(\beta_{sh}+1\right)^4 \Gamma_{sh}^4-1}\pm\sqrt{\dfrac{\left(\beta_{sh}+1\right)^2 \Gamma_{sh}^4-1}{\dfrac{1-\dfrac{1}{\Gamma_{sh}^2}}{1-\dfrac{1}{\left(\beta_{sh}+1\right)^2 \Gamma_{sh}^2}}+1}}\right) \label{eq:beam4veloc}
\end{equation}


\bibliography{sample631}{}

\begin{thebibliography}{}
\expandafter\ifx\csname natexlab\endcsname\relax\def\natexlab#1{#1}\fi
\providecommand{\url}[1]{\href{#1}{#1}}
\providecommand{\dodoi}[1]{doi:~\href{http://doi.org/#1}{\nolinkurl{#1}}}
\providecommand{\doeprint}[1]{\href{http://ascl.net/#1}{\nolinkurl{http://ascl.net/#1}}}
\providecommand{\doarXiv}[1]{\href{https://arxiv.org/abs/#1}{\nolinkurl{https://arxiv.org/abs/#1}}}

\bibitem[{Abbott {et~al.}(2017)Abbott, Abbott, Abbott, \& Acernese}]{Abbott_2017}
Abbott, B.~P., Abbott, R., Abbott, T.~D., \& Acernese, F. 2017, The Astrophysical Journal Letters, 848, L12, \dodoi{10.3847/2041-8213/aa91c9}

\bibitem[{Allen {et~al.}(2012)Allen, Yakimenko, Babzien, Fedurin, Kusche, \& Muggli}]{Allen_2012}
Allen, B., Yakimenko, V., Babzien, M., {et~al.} 2012, Phys. Rev. Lett., 109, 185007, \dodoi{10.1103/PhysRevLett.109.185007}

\bibitem[{Amano \& Hoshino(2009)}]{Amano2009}
Amano, T., \& Hoshino, M. 2009, The Astrophysical Journal, 690, 244, \dodoi{10.1088/0004-637X/690/1/244}

\bibitem[{Begelman \& Kirk(1990)}]{Begelman1990}
Begelman, M.~C., \& Kirk, J.~G. 1990, The Astrophysical Journal, 353, 66, \dodoi{10.1086/168590}

\bibitem[{Bell {et~al.}(2013)Bell, Schure, Reville, \& Giacinti}]{Bell2013}
Bell, A.~R., Schure, K.~M., Reville, B., \& Giacinti, G. 2013, Monthly Notices of the Royal Astronomical Society, 431, 415, \dodoi{10.1093/mnras/stt179}

\bibitem[{Bludman {et~al.}(1960)Bludman, Watson, \& Rosenbluth}]{Bludman1960}
Bludman, S.~A., Watson, K.~M., \& Rosenbluth, M.~N. 1960, The Physics of Fluids, 3, 747, \dodoi{10.1063/1.1706121}

\bibitem[{Bresci {et~al.}(2023)Bresci, Lemoine, \& Gremillet}]{Bresci2023}
Bresci, V., Lemoine, M., \& Gremillet, L. 2023, Physical Review Research, 5, 023194, \dodoi{10.1103/PhysRevResearch.5.023194}

\bibitem[{Bret {et~al.}(2004)Bret, Firpo, \& Deutsch}]{Bret2004}
Bret, A., Firpo, M.-C., \& Deutsch, C. 2004, Physical Review E, 70, 046401, \dodoi{10.1103/PhysRevE.70.046401}

\bibitem[{Bret {et~al.}(2005)Bret, Firpo, \& Deutsch}]{Bret2005}
---. 2005, Physical Review E, 72, 016403, \dodoi{10.1103/PhysRevE.72.016403}

\bibitem[{Bret {et~al.}(2013)Bret, Stockem, Fiuza, Ruyer, Gremillet, Narayan, \& Silva}]{Bret2013}
Bret, A., Stockem, A., Fiuza, F., {et~al.} 2013, Physics of Plasmas, 20, \dodoi{10.1063/1.4798541}

\bibitem[{Buneman(1958)}]{Buneman1958}
Buneman, O. 1958, Physical Review Letters, 1, 8, \dodoi{10.1103/PhysRevLett.1.8}

\bibitem[{{Davidson}(1983)}]{Davidson_1983}
{Davidson}, R.~C. 1983, in Basic Plasma Physics: Selected Chapters, Handbook of Plasma Physics, Volume 1, 229

\bibitem[{Fried(1959)}]{Fried1959}
Fried, B.~D. 1959, The Physics of Fluids, 2, 337, \dodoi{10.1063/1.1705933}

\bibitem[{Galeev \& Sudan(1983)}]{Galeev1983}
Galeev, A., \& Sudan, R.~N. 1983, Basic plasma physics (North-Holland Pub.).
\newblock \url{https://inis.iaea.org/search/search.aspx?orig_q=RN:15068896}

\bibitem[{Gallant \& Achterberg(1999)}]{Gallant1999}
Gallant, Y.~A., \& Achterberg, A. 1999, Monthly Notices of the Royal Astronomical Society, 305, L6, \dodoi{10.1046/j.1365-8711.1999.02566.x}

\bibitem[{Gedalin(1999)}]{Gedalin_99}
Gedalin, M. 1999, Geophysical Research Letters, 26, 1239, \dodoi{https://doi.org/10.1029/1999GL900239}

\bibitem[{Godfrey \& Vay(2015)}]{Godfrey2015}
Godfrey, B.~B., \& Vay, J.-L. 2015, Computer Physics Communications, 196, 221, \dodoi{10.1016/j.cpc.2015.06.008}

\bibitem[{Gottlieb {et~al.}(2018)Gottlieb, Nakar, Piran, \& Hotokezaka}]{Gottlieb2018}
Gottlieb, O., Nakar, E., Piran, T., \& Hotokezaka, K. 2018, Monthly Notices of the Royal Astronomical Society, \dodoi{10.1093/mnras/sty1462}

\bibitem[{Hededal {et~al.}(2004)Hededal, Haugbølle, Frederiksen, \& Nordlund}]{Hededal_2004}
Hededal, C.~B., Haugbølle, T., Frederiksen, J.~T., \& Nordlund, {\AA}. 2004, The Astrophysical Journal, 617, L107–L110, \dodoi{10.1086/427387}

\bibitem[{Hillas(2005)}]{Hillas2005}
Hillas, A.~M. 2005, Journal of Physics G: Nuclear and Particle Physics, 31, R95, \dodoi{10.1088/0954-3899/31/5/R02}

\bibitem[{Ichimaru(1973)}]{ichimaru1973basic}
Ichimaru, S. 1973, Basic Principles of Plasma Physics: A Statistical Approach, Basic Principles of Plasma Physics: A Statistical Approach No. vol.~805387536 (W. A. Benjamin).
\newblock \url{https://books.google.pt/books?id=YiRRAAAAMAAJ}

\bibitem[{Ishihara {et~al.}(1981)Ishihara, Hirose, \& Langdon}]{Ishihara1981}
Ishihara, O., Hirose, A., \& Langdon, A.~B. 1981, The Physics of Fluids, 24, 452, \dodoi{10.1063/1.863392}

\bibitem[{Kasliwal {et~al.}(2017)Kasliwal, Nakar, Singer, Kaplan, Cook, Sistine, Lau, Fremling, Gottlieb, Jencson, Adams, Feindt, Hotokezaka, Ghosh, Perley, Yu, Piran, Allison, Anupama, Balasubramanian, Bannister, Bally, Barnes, Barway, Bellm, Bhalerao, Bhattacharya, Blagorodnova, Bloom, Brady, Cannella, Chatterjee, Cenko, Cobb, Copperwheat, Corsi, De, Dobie, Emery, Evans, Fox, Frail, Frohmaier, Goobar, Hallinan, Harrison, Helou, Hinderer, Ho, Horesh, Ip, Itoh, Kasen, Kim, Kuin, Kupfer, Lynch, Madsen, Mazzali, Miller, Mooley, Murphy, Ngeow, Nichols, Nissanke, Nugent, Ofek, Qi, Quimby, Rosswog, Rusu, Sadler, Schmidt, Sollerman, Steele, Williamson, Xu, Yan, Yatsu, Zhang, \& Zhao}]{Kasliwal2017}
Kasliwal, M.~M., Nakar, E., Singer, L.~P., {et~al.} 2017, Science, 358, 1559, \dodoi{10.1126/science.aap9455}

\bibitem[{Keshet {et~al.}(2009)Keshet, Katz, Spitkovsky, \& Waxman}]{Keshet2009}
Keshet, U., Katz, B., Spitkovsky, A., \& Waxman, E. 2009, The Astrophysical Journal, 693, L127, \dodoi{10.1088/0004-637X/693/2/L127}

\bibitem[{Lampe {et~al.}(1974)Lampe, Haber, Orens, \& Boris}]{Lampe1974}
Lampe, M., Haber, I., Orens, J.~H., \& Boris, J.~P. 1974, The Physics of Fluids, 17, 428, \dodoi{10.1063/1.1694733}

\bibitem[{Lemoine \& Pelletier(2011)}]{Lemoine2011}
Lemoine, M., \& Pelletier, G. 2011, Monthly Notices of the Royal Astronomical Society, 417, 1148, \dodoi{10.1111/j.1365-2966.2011.19331.x}

\bibitem[{Leroy {et~al.}(1981)Leroy, Goodrich, Winske, Wu, \& Papadopoulos}]{Leroy1981}
Leroy, M.~M., Goodrich, C.~C., Winske, D., Wu, C.~S., \& Papadopoulos, K. 1981, Geophysical Research Letters, 8, 1269, \dodoi{10.1029/GL008i012p01269}

\bibitem[{Lyubarsky \& Eichler(2006)}]{Lyubarsky2006}
Lyubarsky, Y., \& Eichler, D. 2006, The Astrophysical Journal, 647, 1250, \dodoi{10.1086/505523}

\bibitem[{Malkov {et~al.}(2024)Malkov, Giacalone, \& Guo}]{malkov2024flat}
Malkov, M., Giacalone, J., \& Guo, F. 2024, Flat Spectra of Energetic Particles in Interplanetary Shock Precursors.
\newblock \doarXiv{2401.07229}

\bibitem[{Medvedev \& Loeb(1999)}]{Medvedev1999}
Medvedev, M.~V., \& Loeb, A. 1999, The Astrophysical Journal, 526, 697, \dodoi{10.1086/308038}

\bibitem[{Nakar {et~al.}(2011)Nakar, Bret, \& Milosavljević}]{Nakar_2011}
Nakar, E., Bret, A., \& Milosavljević, M. 2011, The Astrophysical Journal, 738, 93, \dodoi{10.1088/0004-637X/738/1/93}

\bibitem[{Niemiec \& Ostrowski(2004)}]{Niemiec2004}
Niemiec, J., \& Ostrowski, M. 2004, The Astrophysical Journal, 610, 851, \dodoi{10.1086/421730}

\bibitem[{Ohira \& Takahara(2008)}]{Ohira2008}
Ohira, Y., \& Takahara, F. 2008, The Astrophysical Journal, 688, 320, \dodoi{10.1086/592182}

\bibitem[{Papadopoulos {et~al.}(1971)Papadopoulos, Davidson, Dawson, Haber, Hammer, Krall, \& Shanny}]{Papadopoulos1971}
Papadopoulos, K., Davidson, R.~C., Dawson, J.~M., {et~al.} 1971, The Physics of Fluids, 14, 849, \dodoi{10.1063/1.1693520}

\bibitem[{Paschmann {et~al.}(1980)Paschmann, Sckopke, Asbridge, Bame, \& Gosling}]{Paschmann1980}
Paschmann, G., Sckopke, N., Asbridge, J., Bame, S., \& Gosling, J. 1980, Journal of Geophysical Research: Space Physics, 85, 4689, \dodoi{10.1029/JA085iA09p04689}

\bibitem[{Piran(2005)}]{Piran2005}
Piran, T. 2005, Reviews of Modern Physics, 76, 1143, \dodoi{10.1103/RevModPhys.76.1143}

\bibitem[{Plotnikov {et~al.}(2013)Plotnikov, Pelletier, \& Lemoine}]{Plotnikov2013}
Plotnikov, I., Pelletier, G., \& Lemoine, M. 2013, Monthly Notices of the Royal Astronomical Society, 430, 1280, \dodoi{10.1093/mnras/sts696}

\bibitem[{Rabinak {et~al.}(2011)Rabinak, Katz, \& Waxman}]{Rabinak2011}
Rabinak, I., Katz, B., \& Waxman, E. 2011, The Astrophysical Journal, 736, 157, \dodoi{10.1088/0004-637X/736/2/157}

\bibitem[{Rybicki \& Lightman(1985)}]{Rybicki1985}
Rybicki, G.~B., \& Lightman, A.~P. 1985, Radiative Processes in Astrophysics (Wiley), \dodoi{10.1002/9783527618170}

\bibitem[{Shimada \& Hoshino(2000)}]{Shimada2000}
Shimada, N., \& Hoshino, M. 2000, The Astrophysical Journal, 543, L67, \dodoi{10.1086/318161}

\bibitem[{Silva {et~al.}(2003)Silva, Fonseca, Tonge, Dawson, Mori, \& Medvedev}]{Silva_2003}
Silva, L.~O., Fonseca, R.~A., Tonge, J.~W., {et~al.} 2003, The Astrophysical Journal, 596, L121, \dodoi{10.1086/379156}

\bibitem[{Sironi \& Spitkovsky(2011)}]{Sironi2011}
Sironi, L., \& Spitkovsky, A. 2011, The Astrophysical Journal, 726, 75, \dodoi{10.1088/0004-637X/726/2/75}

\bibitem[{Smartt {et~al.}(2017)Smartt, Chen, Jerkstrand, Coughlin, Kankare, Sim, Fraser, Inserra, Maguire, Chambers, Huber, Krühler, Leloudas, Magee, Shingles, Smith, Young, Tonry, Kotak, Gal-Yam, Lyman, Homan, Agliozzo, Anderson, Angus, Ashall, Barbarino, Bauer, Berton, Botticella, Bulla, Bulger, Cannizzaro, Cano, Cartier, Cikota, Clark, Cia, Valle, Denneau, Dennefeld, Dessart, Dimitriadis, Elias-Rosa, Firth, Flewelling, Flörs, Franckowiak, Frohmaier, Galbany, González-Gaitán, Greiner, Gromadzki, Guelbenzu, Gutiérrez, Hamanowicz, Hanlon, Harmanen, Heintz, Heinze, Hernandez, Hodgkin, Hook, Izzo, James, Jonker, Kerzendorf, Klose, Kostrzewa-Rutkowska, Kowalski, Kromer, Kuncarayakti, Lawrence, Lowe, Magnier, Manulis, Martin-Carrillo, Mattila, McBrien, Müller, Nordin, O’Neill, Onori, Palmerio, Pastorello, Patat, Pignata, Podsiadlowski, Pumo, Prentice, Rau, Razza, Rest, Reynolds, Roy, Ruiter, Rybicki, Salmon, Schady, Schultz, Schweyer, Seitenzahl, Smith, Sollerman, Stalder, Stubbs, Sullivan, Szegedi,
  Taddia, Taubenberger, Terreran, van Soelen, Vos, Wainscoat, Walton, Waters, Weiland, Willman, Wiseman, Wright, Wyrzykowski, \& Yaron}]{Smartt2017}
Smartt, S.~J., Chen, T.-W., Jerkstrand, A., {et~al.} 2017, Nature, 551, 75, \dodoi{10.1038/nature24303}

\bibitem[{Vanthieghem {et~al.}(2022)Vanthieghem, Lemoine, \& Gremillet}]{Vanthieghem2022}
Vanthieghem, A., Lemoine, M., \& Gremillet, L. 2022, The Astrophysical Journal Letters, 930, L8, \dodoi{10.3847/2041-8213/ac634f}

\bibitem[{Weibel(1959)}]{Weibel1959}
Weibel, E.~S. 1959, Physical Review Letters, 2, 83, \dodoi{10.1103/PhysRevLett.2.83}

\bibitem[{Yoon \& Davidson(1987)}]{Yoon_87}
Yoon, P.~H., \& Davidson, R.~C. 1987, Phys. Rev. A, 35, 2718, \dodoi{10.1103/PhysRevA.35.2718}

\bibitem[{Zank {et~al.}(2000)Zank, Rice, \& Wu}]{Zank2000}
Zank, G.~P., Rice, W. K.~M., \& Wu, C.~C. 2000, Journal of Geophysical Research: Space Physics, 105, 25079, \dodoi{10.1029/1999JA000455}

\end{thebibliography}
\bibliographystyle{aasjournal}



\end{document}